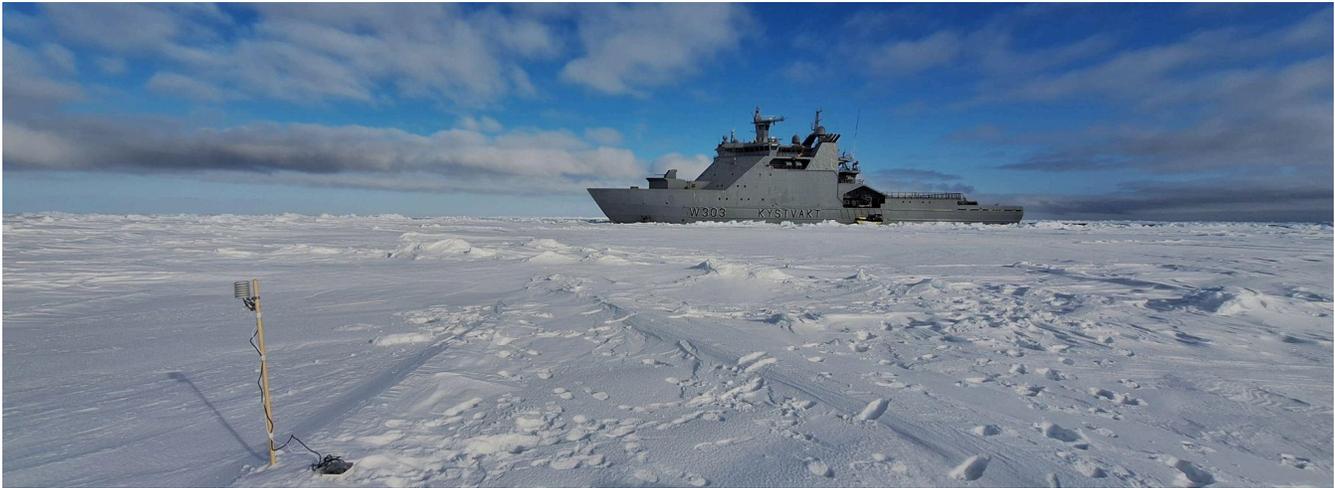
Picture credit (J. Tjernström)

# Cruise Report - KV Svalbard 25.April-11.May 2025

## SvalMIZ-25: Svalbard Marginal Ice Zone 2025 Campaign

M. Müller[1,2,*], J. Rabault[1], C. Palerme[1] and J. Tjernström[1,2]

(1) Norwegian Meteorological Institute
(2) Department of Geosciences, University of Oslo
(*) corresponding author: maltem@met.no

# 1. Overview

The coupling of weather, sea-ice, ocean, and wave forecasting systems has been a long-standing research focus to improve Arctic forecasting systems and their realism and is also a priority of international initiatives such as the WMO research project PCAPS. The goal of the **Svalbard Marginal Ice Zone 2025 Campaign (SvalMIZ-25)** was to observe and better understand the complex interplay between atmosphere, waves, and sea-ice in the winter Marginal Ice Zone (MIZ) in order to advance predictive skill of coupled Arctic forecasting systems. The main objective has been to set up a network of observations with a spatial distribution that allows for a representative comparison between in situ observations and gridded model data. The observed variables include air and surface temperature, sea-ice drift, and wave energy spectra.

With the support of the Norwegian Coast Guard, we participated in the research cruise with KV Svalbard from 22.April - 11.May 2025. In total 21 buoys were deployed in the Marginal Ice Zone north of the Svalbard Archipelago. The first part of the report describes the instruments and their calibration (Section 2), and the second part briefly describes the weather, sea ice, and wave conditions during the campaign. The quality-controlled and calibrated data will soon be made publicly available as part of a scientific data paper and as a Zenodo data release[1] under the doi **10.5281/zenodo.17087019**. Note that the first version of the data that is published is not yet quality controlled.

## 2. Instruments/Measurements

### a. OpenMetBuoy: Wave, temperature, and infrared sensors

The OpenMetBuoy (OMB, *Rabault et al. 2021*) is used as the basis for the instruments deployed. The OMB is an open source, low cost (in its standard configuration: around 650 USD construction cost all included, and 110 USD per month of activity for iridium cost) buoy with global communication capability through the iridium short burst data global network. The original OMB performs measurements of drift (using GPS, at a sample interval of 30 minutes) and 1-dimensional wave spectrum (using motion sensors, embedded Kalman filtering, and in-situ signal processing; the raw data are sampled at 800Hz for 20 minutes, and a new spectrum is acquired every 2 hours by default). More details are available, e.g., in *Rabault et. al. 2022*, or in previous datasets using the OMB, e.g. *Rabault et. al. 2023*. The OMB GPS and wave measurements are well-validated in the literature. The OMB is adapted to performing measurements in the polar regions and can function for up to 4.5 months in polar temperatures using just 2 non-rechargeable lithium D-cells.

For the *SvalMIZ-25 Campaign*, the OMB has been extended by three temperature sensors and one infrared sensor pointing downward to the snow surface:

The three temperature sensors are as follows:
1. One temperature sensor is mounted at 1-meter height within a radiation shield.
2. One temperature sensor is positioned at the snow-ice interface.
3. One temperature sensor is positioned at 0.3 meter depth in the sea-ice.

The IR sensor (MLX90614) has been mounted below the radiation shield at 1-meter height pointing downward to the snow surface, measuring the snow surface temperature. The general setup is illustrated in Figure 1.

---

[1] https://zenodo.org/records/17087019

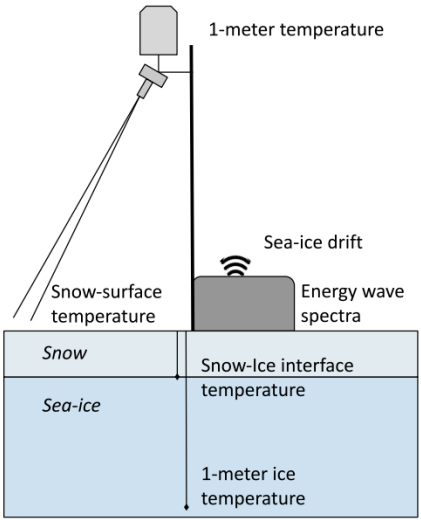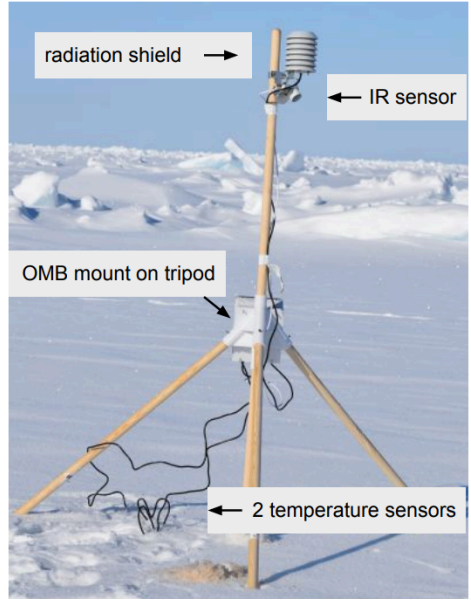

**Figure 1:** The SvalMIZ-25 setup with OMB including the GPS for position measurement and IMU for wave measurements, the 1-meter pole with a temperature sensor within a radiation shield and an IR sensor, as well as, the two temperature sensors measuring at the snow-ice interface and 0.3 meter in the ice.

The three temperature sensors are DS18B20 digital sensors with 1-wire bus interface. The temperature sensors are individually calibrated. For this, a temperature-controlled calibration cabinet at the Norwegian Meteorological Institute is used to perform reference temperature readings at 3 set temperatures (-7.5 degrees C, 0 degrees C, +5 degrees C). A reference PT-100 sensor is used to obtain the exact temperature in the cabinet, and this is used to perform a 3-point calibration of each DS18B20 individually. We find that the sensors follow, to a very good approximation, a linear calibration curve. Therefore, we apply a linear calibration to each DS18B20 individually. This is similar to the methodology that was applied in *Müller et al. 2024, 2025a, 2025b*.

In addition to the calibration of the sensors per se, experimental setups have been conducted in the Oslo test field of the Norwegian Meteorological Institute to understand the sensitivities to the radiation shield used for the 1-meter sensor. For testing of the radiation shield (Rikasensor RK95-02B), we used a reference laboratory-quality PT-100 sensor in a standard professional-grade radiation shield and compared the temperature values obtained by the PT100 in the professional-grade radiation shield, to one calibrated DS18B20 in the radiation shield used for SvalMIZ-25. We find that the new radiation shields are significantly more effective than the ones used in *Müller et al. 2025b*, so that we expect no significant solar radiation bias in the SvalMIZ-25 dataset.

The IR sensor used is the MLX90614 DCC sensor module[2]. The field of view of the sensor is 35°, and according to the manufacturer, the temperature measurement range is -70°C～380°C, and the working temperature is between -40°~85°C. In order to cut off the visible and near infra-red radiant flux, which otherwise distort the temperature measurements by the sensor, a version of the MLX90614 that includes an optical filter (long-wave bandpass) is used. The wavelength pass band of this optical filter is from 5.5 to 14µm.

---

[2] https://www.seeedstudio.com/Grove-Thermal-Imaging-Camera-MLX90614-DCC-IR-Array-with-35-FOV-p-4657.html

We evaluated the IR sensor against a reference professional-grade IR sensor from Campbell Science (IR120 sensor). This sensor has a field of view of 40° and was mounted next to the MLX90614 sensor, pointing down at a snow surface at a test field in Oslo in the winter of 2024/2025. The results show that the MLZ90614 sensor is noisier than the reference IR120 sensor when measuring with a sampling rate of 0.1Hz (standard deviation of about 0.4°C), see Figure 2. The mean deviation between the two sensors is 0.16°C. However, generally, the variability in temperatures between -15°C and 0°C is captured well, and a good agreement as well as low noise, is obtained by averaging readings from the MLX90614. On the OMB, 60 readings by the MKX90614 sensor taken over a period of 60 seconds are averaged to obtain the measurement that is transmitted over Iridium.

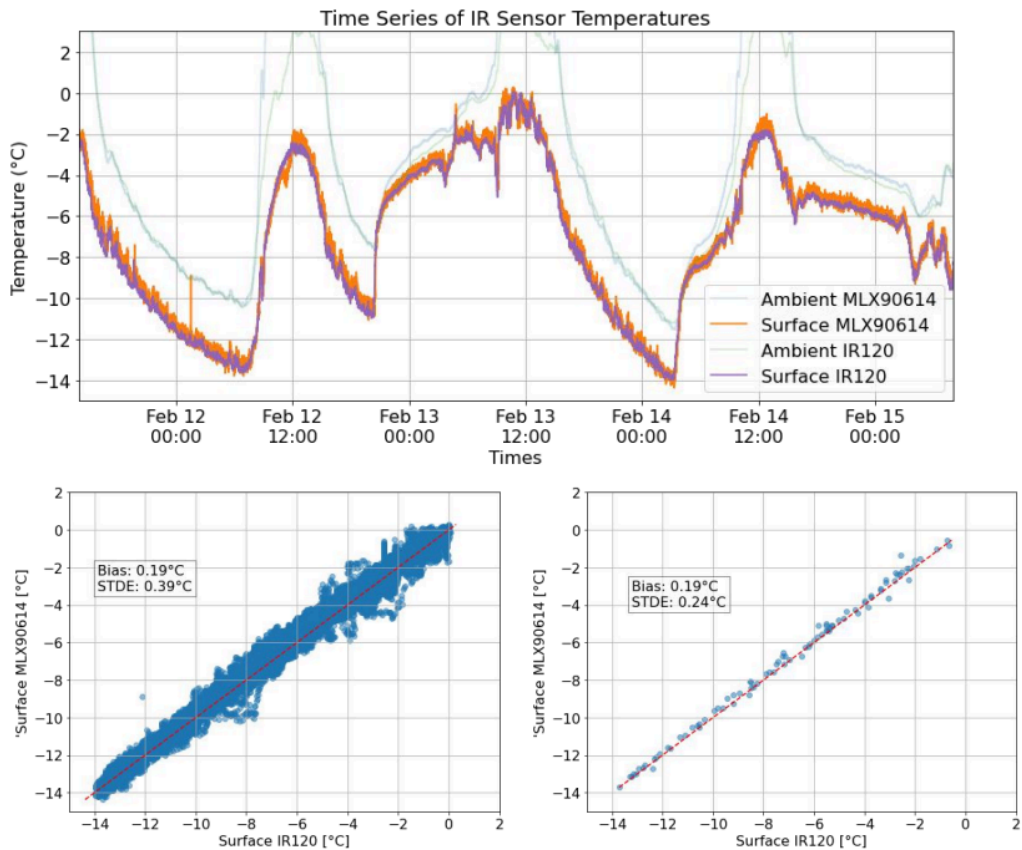

**Figure 2:** Top: Time series of the MLX90614 and IR120 temperature measurements (snow surface and ambient), with 10 and 5 second sampling rates, respectively. Bottom left: Scatter plot of 0.1Hz snow surface temperature of IR120 vs MLX90614. Bottom right: Scatter plot of hourly-averaged temperature samples.

In addition to the buoys described above, one prototype of buoy equipped with a temperature measurement string was deployed (buoy KVS-17). This buoy is a standard OMB, equipped with the same 1m air temperature measurement and IR downward-looking sensor as described above. However, instead of the two extra DS18B20 sensors measuring the snow-ice and 30cm in-ice temperatures, KVS-17 is equipped with a temperature string that is frozen into the ice. The temperature string is equipped with 26 DS18B20 sensors, which are calibrated similarly to described above. The temperature string is 4m long, and the 26 temperature sensors are equally distributed along the string. Since it uses a different data format, the dataset for this buoy is released as an independent netCDF file. More details are available in the metadata of the file, see the online data release.

### b. Ice thickness and snow depth measurements

Sea-ice thickness observations were collected at 3 locations close by (ca. 5-10 meters) to the OpenMetBuoy station. For each station, we collected 40 snow depth measurements along a square with an avalanche probe providing a 1 cm snow depth measurement resolution. The size of the square depends on sea ice floe characteristics (size, ridges) on which the stations were deployed, and is typically between 30 and 60 meters on each side.

### c. OVL Portal

The OceanVirtualLaboratory (OVL, developed by OceanDatalab, Locmaria-Plouzané, France; see Collard et. al. 2015) has been used to monitor the data and to plan the buoy deployment sites during the campaign (see Figure 3). The uncalibrated temperature data and the wave spectra can be assessed via the permalink https://odl.bzh/2_B1zdSE and can be visualized interactively in combination with various sources of satellite and model output data. Data for this campaign can be found under *Products/In Situ/ OpenMetBuoys Svalbard MIZ*.

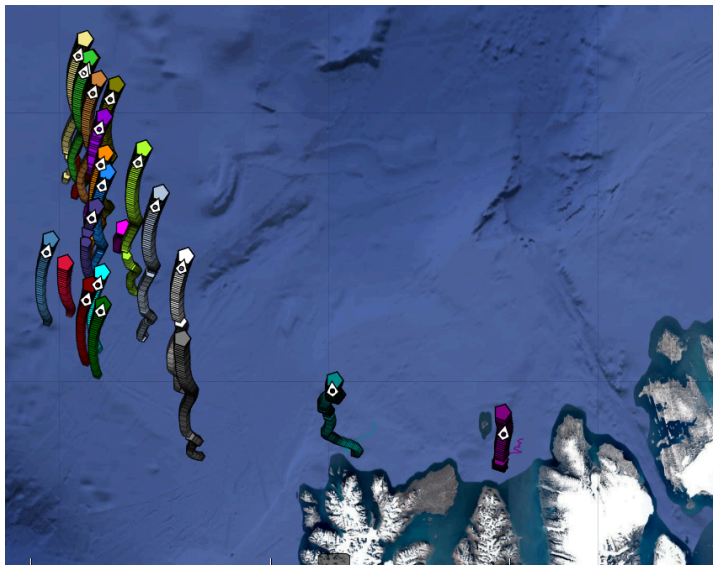

**Figure 3**: Illustration of the OVL web-based visualization of the buoys trajectories. To interactively browse the data, see https://odl.bzh/2_B1zdSE

# 3. Campaign, results, and data location

## a. Overview

In total, 21 OceanMetBuoys have been successfully deployed during the cruise on the sea-ice North-west of Svalbard. At the time of deployment, a significant amount of sea ice was present in the area, as visible in Figure 4.

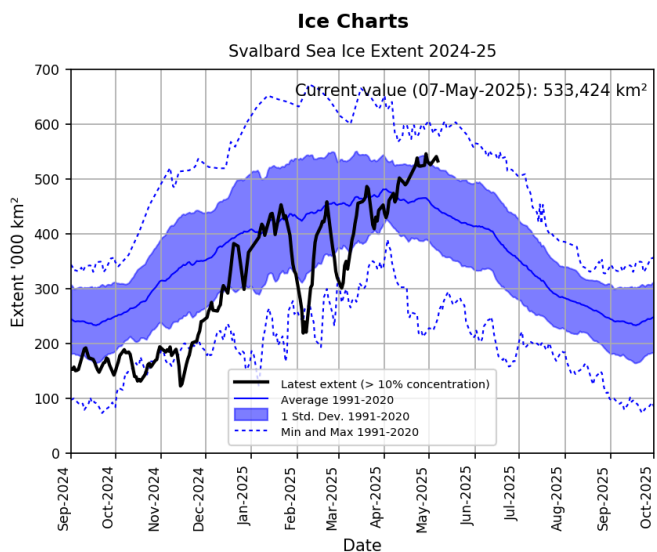

**Figure 4:** Climatology of sea-ice extent around Svalbard and values of the 6 months preceding the cruise. Original figure from cryo.met.no.

## b. Weather, wave, and sea-ice conditions

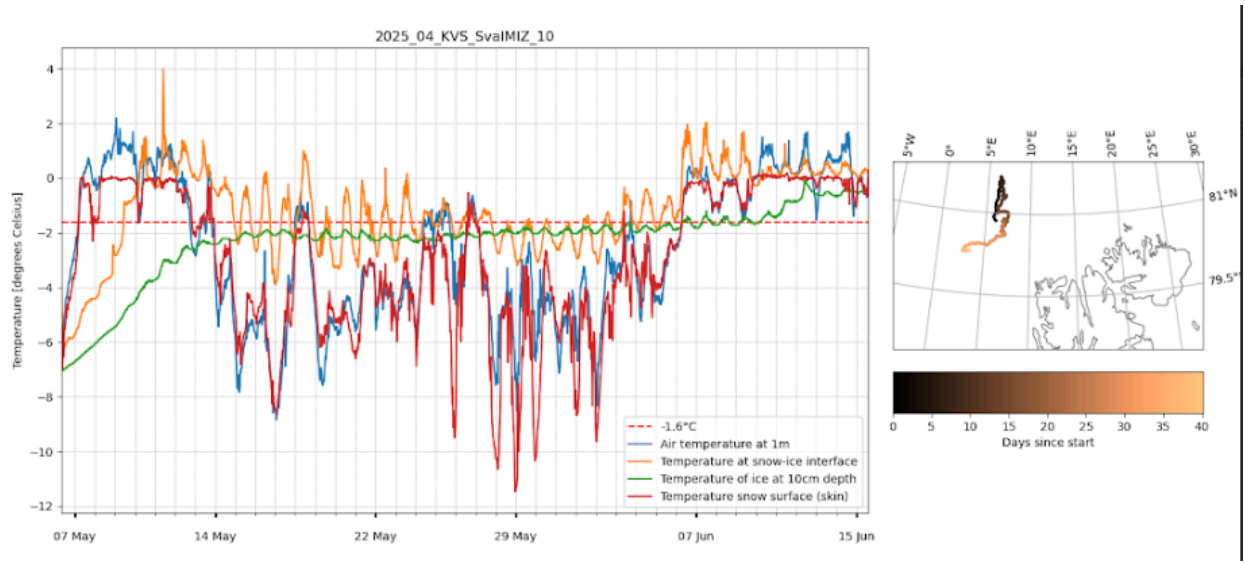

**Figure 5**: 1m-temperature, snow-ice, 30cm in ice and skin temperature measurements of buoy KVS-10.

During the first days of the deployment in early May, a storm system advected warm air northward across the buoy transect. This was accompanied by large ocean waves propagating along the array, and in addition was leading to enhanced variability in both air and snow surface temperatures. After this event, conditions shifted to a sustained cold phase lasting until the end of May, with air and snow surface temperatures remaining below freezing and dropping as low as −10 °C. From around 7 June onward, a marked warming occurred, with air and surface temperatures persistently rising above 0 °C, indicating the onset of melt conditions. This is illustrated by Figure 5.

## c. Ice thickness and snow depth.

The sea-ice thickness and snow depth mean and standard deviation (numerical values available in Table 1) from all stations are shown in FIg. 6.

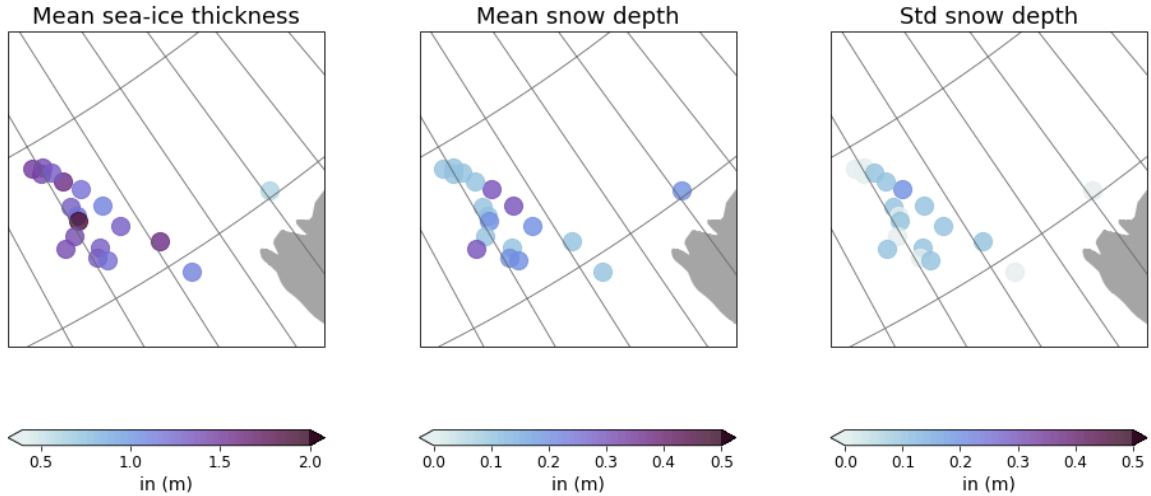

**Figure 6:** The mean sea-ice thickness, mean snow depth (see Table 1), and standard deviation (std) of the snow depth observed during the deployment of the buoys from April 27 to May 7, 2025.

### d. Data availability

Data and scripts used to generate these will be made available on Github [3] as a netCDF format file following the CF conventions, as provided by the TraJan [4] project, once they are quality controlled. In the meantime, data can be requested directly from the authors. In addition, an archive of versioned datasets is available on Zenodo [5].

---

[3] https://github.com/jerabaul29/2025_Svalbard_MIZ_KVS_SvalMIZ25
[4] https://github.com/OpenDrift/trajan
[5] https://zenodo.org/records/17087019

SvalMIZ-25: Svalbard Marginal Ice Zone Campaign 2025

# Appendix

| ID | Date | Time | Mean Ice Thickness (m) | Mean Snow Depth (m) | Std Snow Depth (m) |
|---|---|---|---|---|---|
| 2025_04_KVS_SvalMIZ_24 | 27.4.2025 | 9:00:00 AM | 0.4 | 0.1 | 0.0 |
| 2025_04_KVS_SvalMIZ_06 | 2.5.2025 | 4:30:00 PM | 0.6 | 0.2 | 0.0 |
| 2025_04_KVS_SvalMIZ_01 | 5.5.2025 | 8:19:00 AM | 1.1 | 0.1 | 0.0 |
| 2025_04_KVS_SvalMIZ_09 | 5.5.2025 | 10:32:00 AM | 1.6 | 0.1 | 0.1 |
| 2025_04_KVS_SvalMIZ_22 | 5.5.2025 | 11:45:00 AM | 1.3 | 0.2 | 0.1 |
| 2025_04_KVS_SvalMIZ_08 | 5.5.2025 | 2:21:00 PM | 1.1 | 0.3 | 0.1 |
| 2025_04_KVS_SvalMIZ_23 | 5.5.2025 | 4:00:00 PM | 1.2 | 0.3 | 0.2 |
| 2025_04_KVS_SvalMIZ_04 | 5.5.2025 | 5:53:00 PM | 1.6 | 0.1 | 0.1 |
| 2025_04_KVS_SvalMIZ_03 | 5.5.2025 | 8:25:00 PM | 1.4 | 0.1 | 0.0 |
| 2025_04_KVS_SvalMIZ_25 | 5.5.2025 | 10:23:00 PM | 1.5 | 0.1 | 0.0 |
| 2025_04_KVS_SvalMIZ_10 | 6.5.2025 | 8:11:00 AM | 1.4 | 0.1 | 0.0 |
| 2025_04_KVS_SvalMIZ_07 | 6.5.2025 | 12:44:00 PM | 1.3 | 0.1 | 0.1 |
| 2025_04_KVS_SvalMIZ_17 | 6.5.2025 | 4:00:00 PM | 2.0 | 0.1 | 0.0 |
| 2025_04_KVS_SvalMIZ_15 | 6.5.2025 | 8:26:00 PM | 1.3 | 0.1 | 0.1 |
| 2025_04_KVS_SvalMIZ_18 | 6.5.2025 | 10:11:00 PM | 1.2 | 0.1 | 0.0 |
| 2025_04_KVS_SvalMIZ_05 | 7.5.2025 | 8:36:00 AM | 1.9 | 0.2 | 0.1 |
| 2025_04_KVS_SvalMIZ_11 | 7.5.2025 | 12:18:00 PM | 1.4 | 0.1 | 0.0 |
| 2025_04_KVS_SvalMIZ_20 | 7.5.2025 | 1:40:00 PM | 1.4 | 0.3 | 0.1 |
| 2025_04_KVS_SvalMIZ_19 | 7.5.2025 | 4:14:00 PM | 1.3 | 0.1 | 0.1 |
| 2025_04_KVS_SvalMIZ_02 | 7.5.2025 | 7:00:00 PM | 1.3 | 0.2 | 0.0 |
| 2025_04_KVS_SvalMIZ_14 | 7.5.2025 | 8:00:00 PM | 1.2 | 0.2 | 0.1 |

**Table 1:** Overview of the deployment of sensors. **ID** - name of the buoy, **Date and time** of deployment are given in UTC. **Mean and standard deviation of the snow depth** - calculated from up to 20 measurements in an area of 20-40 meters around the sensor. **Mean Ice thickness** - calculated from three measurements from about 5 meters distance to the pole.

# References

Collard, F., Quartly, G., Konik, M., Johannessen, J., Korosov, A., Chapron, B., Piollé, J., Herlédan, S., Darecki, M., Isar, A., et al.: Ocean Virtual Laboratory: A new way to explore multi-sensor synergy demonstrated over the Agulhas region, in: Proceedings of Sentinel-3 for Science Workshop (2-5 June 2015, Venice, Italy), ESA, 2015.

Müller, Malte, Jean Rabault, and Cyril Palerme. "Svalbard Marginal Ice Zone 2024 Campaign--Cruise Report." *arXiv preprint arXiv:2407.18936* (2024).

Müller, Malte, et al. "Distributed observation networks in the Arctic Marginal Ice Zone to advance forecasting systems." *Bulletin of the American Meteorological Society* 106.6 (2025a): E1204-E1210.

Malte Müller, et al. "Svalbard Marginal Ice Zone 2024: A distributed network of temperature, waves, and sea ice drift observations" Scientific Data (2025b)

Rabault, Jean, et al. "Openmetbuoy-v2021: An easy-to-build, affordable, customizable, open-source instrument for oceanographic measurements of drift and waves in sea ice and the open ocean." *Geosciences* 12.3 (2022): 110.

Rabault, Jean, et al. "A dataset of direct observations of sea ice drift and waves in ice." *Scientific Data* 10.1 (2023): 251.

# Acknowledgment

This campaign has been supported by the WMO PCAPS projects. We would like to acknowledge the great support we had from across the Norwegian Meteorological Institute. In particular, Bikas C. Bhattarai and Olaf Weisser for supporting us with the temperature sensor calibration. Gaute Hope for integrating the buoys into the wavebug system. Steinar Eastwood for loaning us the infrared sensor. Jørn Kristiansen and Audun Christoffersen for supporting our ideas and ambitions. We also thank Marius Jonassen from UNIS for their great help with all our logistics issues. In addition, the support from the OVL team (Fabrice Collar / DrFab, Sylvain Herlédan) and the real-time integration of the buoys into the portal has helped us greatly with the planning during the buoy deployment and monitoring the data. The collaboration and support of the KV Svalbard crew have been central to achieving our research objectives. Their expertise and dedication were crucial in navigating through challenging sea conditions and ensuring the safety and effectiveness of our scientific deployments.